\documentclass[10pt]{article}
\usepackage{graphicx}

\usepackage{epsfig}
\usepackage{latexsym}
\usepackage{amsmath}
\usepackage{eurosym}

\begin{document}

\pdfoutput=1

\title{Exploring how innovation strategies at time of
crisis influence performance: \\ a cluster analysis perspective}

\author{ Marcel Ausloos\textsuperscript{a,b}, \\ 
 Francesca Bartolacci\textsuperscript{c}, \\
 Nicola G. Castellano\textsuperscript{d}, \\
 Roy Cerqueti\textsuperscript{e}\\ \\ 
 $^{a}$ School of
Business, University of Leicester. \\University Road, Leicester, LE1
7RH, United Kingdom.\\ Email: ma683@le.ac.uk \\$^{b}$
GRAPES. rue de la Belle Jardiniere, 483/0021,
\\B-4031, Liege Angleur, Belgium, Euroland. \\Email:
marcel.ausloos@ulg.ac.be 
\\ $^{c,d,e}$Department of Economics and Law, \\University of
Macerata. Via Crescimbeni, 20, I-62100, Macerata, Italy
 \\$^{c}$Email:  bartolacci@unimc.it\\
  $^{d}$Email:
ncaste@unimc.it\\ 
   $^{e}$Email:  roy.cerqueti@unimc.it}
\maketitle 
\newpage
\begin{abstract}

This paper analyzes the connection between innovation activities of
companies -- implemented before crisis -- and their performance --
measured at time of crisis. The companies listed in the STAR Market
Segment of the Italian Stock Exchange are analyzed. Innovation is
measured through the level of investments in total tangible and
intangible fixed assets in 2006-2007, while performance is captured
through growth -- expressed by variations of sales, total assets and
employees -- profitability -- through ROI or ROS -- and productivity
-- through asset turnover or sales per employee in the period
2008-2010. The variables of interest are analyzed and compared
through statistical techniques and by adopting cluster analysis. In
particular, a Voronoi tessellation is also implemented in a varying
centroids framework. In accord with a large part of the literature,
we find that the behavior of the performance of the companies is not
univocal when they innovate.
\end{abstract}

\vskip0.2cm


\section{Introduction}

 The efforts spent by the entrepreneurs in innovation
initiatives have the specific target of contributing to enhance
companies performances. However, the real effect of innovation on
performance is still at the center of the academic debate, also for
the practical implications of this theme. Furthermore, the influence
of the status of the economic environment is also of paramount
relevance for the complete understanding of such a relationship. 
\newline
The present research aims  at exploring the connection
between innovation strategy and performance of a company at time of
a global  economic crisis.  In our specific context,
innovation is derived from empirical data and is captured by the
level of investments in innovation activities. Thus, we will split
such a conceptualization into tangible and intangible assets, as we
will see below. {In our study, the relationship between
various innovation initiatives and performance measures of Italian
companies is investigated from quantitative information available in
their publicized consolidated financial statements. Specifically, we
deal with the whole set of companies listed in the STAR market
segment of Italian Stock Exchange. 
\newline
Such a market includes only "mid-sized companies" in terms of
capitalization. Hence, the investigated sample is coherent with a
common sense aim that the impacts produced by innovation on
performance should be more evident in companies focusing on a unique
or on rather limited fields of operating activity. Moreover, the
sample of such listed companies allows an exhaustive
availability of financial statements information, preventing any
bias due to companies selection. 
.
\newline
 We have assumed that the generation of innovation requires
significant investment of resources (Heirman and Clarysse, 2007;
Renzi and Simone, 2011; Montresor and Vezzani, 2016) and have
detected in 2006-2007 the presence of innovation initiatives
represented by investment level on intangible and tangible fixed
assets (OECD 2005, p. 35). Indeed, even if these measures do not
cover all aspects of innovation, they surely represent a significant
part of it (see OECD 2005, p. 40). Furthermore the employment of
financial measures to quantify innovation is widely accepted (see
e.g. Chun et al. 2015; Gocer et al. 2016; Baum et al., 2017;
Ceptureanu et al., 2017). We follow such a line of thought. In the
2008-2010 post-crisis time interval, the performance outcomes are
measured through (i) growth (or decay) variations: those of sales,
total assets and number of employees; (ii) profitability: return on
investment (ROI) and return on sales (ROS); (iii) efficiency: assets
turnover and sales per employee. The time-horizon is consistent with
the good practice of productivity in analyzing the effects of
innovation on performance, as clearly declared by OECD 2005.
 Specifically, we take into account aggregated indicators for
growth, profitability and productivity.
\newline
Bartolacci et al. (2015) employ the same sample analyzed here and
discuss the effects of innovation on performance through a new class
of entropy measures. In particular, the quoted paper seeks the
similarities between companies in terms of the disorder generated by
their classifications.
\newline
Also the present study is based on a cluster analysis of firm-level
data -- the STAR market companies are not homogenous neither in
terms of industry nor for the propensity to innovate. Nevertheless,
the heterogeneity can be viewed as a bonus allowing for a credible
focus due to some independence of market constraints specific to a
industry segment.
\newline
However, we differ from Bartolacci et al. (2015) since we propose a
formal and rather original method, based on Voronoi tessellation
(Voronoi, 1908). Such a statistical tool consists of the a-priori
definition of some reference points - namely, centroids -- and of a
distance measure, and each centroid identifies a cluster whose
elements have distance smaller to it than to the other centroids.
\newline
We depart from the original formulation of Voronoi 
by introducing a concept of weighted Euclidean distance, hence
leading to asymmetry (see formulas (\ref{dalpha}) and
(\ref{dbeta})). In so doing, we specify different relative relevance
to the variables, hence gaining insights from the analysis.
\newline
Within the clusters, one could compare the characteristics and
performance of companies holding the same innovation level, whereas
between the clusters heterogeneity means that different innovation
levels might be suitable for obtaining different levels of
performance. In this respect, cluster analysis seems to be
particularly effective in providing a global analysis of the
relationship between innovation and performance but also a
disaggregated discussion of the single units and of the clusters.
\newline
Due to its versatility, the proposed methodology has been applied in
several scientific fields, like neuroscience (see e.g. Duyckaerts
and Godefroy, 2000), astrophysics (see e.g. Ramella et al., 2001)
and material science (see e.g. Gadomski and Kruszewska, 2012).
\newline
However, the use of Voronoi tessellation is quite neglected in the
management literature. Applications of this technique to economic
themes can be found in Liu et al. (2009), Yushimito et al. (2012)
and Vaz et al. (2014). Hence, this paper contributes to fill the gap
between complex science and management.
\newline
More generally, the paper is in line with a large strand of economic
literature. Indeed, clustering techniques are largely employed to
analyse the performance at country, industrial district or firm
level (see e.g. Zahra and Covin, 1994; Gligor and Ausloos, 2007,
2008a, 2008b). Furthermore, in some cases the cluster analysis is
employed to investigate the determinants of innovation and
innovation-performance focused on a single industry (Tseng et al.,
2008), a single country (Dwyer and Mellor, 1993; Vaz et al., 2014;
Agostini et al., 2015), or different industries and/or countries
(Pavitt, 1984; Cesaratto and Mangano, 1993; Leiponen and Drejer,
2007).
\newline
The paper is structured as follows: Section 2 contains a summary of
extant literature about innovation and performance and the research
questions developed accordingly; Section 3 describes the dataset and
the explanatory variables; Section 4 outlines the employed
methodology; Section 5 states and discusses the main results. Last
section concludes.}

\section{Innovation and performance: literature review and research question}
 Scholars generally investigate the characteristics
associated to innovative companies, or the relationships expected
between innovation and performance. In several cases the analyses
are combined. Such studies are of classificatory and predictive
kind.
\newline
In the first stream of research Khan and Manopichetwattana (1989)
investigate innovativeness in small and medium companies, in
association to a set of characteristics (environmental,
organizational, entrepreneurial, etc.). They find that the
inclination to collect information about the external environment is
positively associated with innovation. Similarly, positive impacts
on innovation are produced
by environmental dynamism or heterogeneity. 
\newline
The pioneering paper of Pavitt (1984) classifies companies according
to their innovative activities by using an inductive methodological
approach (Archibugi, 2001). Pavitt''s taxonomy, over time, inspired
numerous scholars applying cluster analysis with the aim to classify
firms according to how intensively they innovate in order to
investigate the effects on performance (Cesaratto and Mangano, 1993;
Dwyer and Mellor, 1993; Hollenstein, 2003; De Jong and Marsili,
2006; Leiponen and Drejer, 2007; Jensen et al., 2007) In many cases,
these works are based on innovation and performance firms data and
they consistently show a heterogeneous behaviour among companies.
\newline
Cooper (1984) adopts a cluster analysis to measure the performance
achieved by product innovation and identify the strategies leading
to different types of performance. Obtained results show that new
product performance is largely decided by the policy that top
management elects in a specific context.
\newline
Hollenstein (2003), studying innovation modes in the Swiss service
sector, finds an unclear association between innovation intensity
and performance, probably due to the impact produced by other
significant determinants of performance. Dwyer and Mellor (1993)
come to similar results after studying five alternative strategic
approaches to product innovation which produce a similar rate of
success and profitability. Leiponen and Drejer (2007) highlight how
different industries behave in terms of company innovation; this
suggests that the characterizing features of the firms, like
strategic behaviors or local search activities, are associated to a
technological framework leading to short term performance
inhomogeneity. In this respect, Srholec and Verspagen (2012) show
that heterogeneity of firms at a sectoral and country level is the
key to understand why companies behave differently when they
innovate. Shin et al (2017) in their recent study find that the
effect produced by innovation, measured in terms of $R\& D$
intensity, on performance is influenced by the level of vertical
integration: in particular, less integrated companies may focus on a
limited part of the innovation process thereby increasing their
profitability compared to more integrated companies of the same
industry.
\newline
Performance heterogeneity may be related to technological
innovation, as shown by several works in reference to many
industrialized countries and industries (Brusoni et al., 2006;
Lawless and Anderson, 1996; Kirner et al., 2009). Park et al. (2012)
introduce a strategic dimension in order to explain heterogeneous
performance in innovative companies, and they find that
technology-oriented companies - compared to market oriented ones -
are more likely to achieve instant performances since, their
strategic focus on the products and services requested by customers
allows managers to strengthen the customer loyalty.
\newline
At country level, Sterlacchini and Venturini (2014) compare Italian
and Spanish companies over a long time period (1980-2006) and find
that $R\& D$ is a crucial driver of manufacturing productivity, and
this is also supported in countries generally classified as
technology followers.
\newline
Cesaratto and Mangano (1993) highlight the variety of behaviours
among and within sectors, and so the absence of a balance between
the resources invested in innovation, quantity and quality of
innovative output and economic financial performance.
\newline
Among the variables affecting the relation between innovation and
performance, the presence of distress of the economic environment
may play a significant role (Ranga and Etzkowitz, 2012; Nunes and
Lopes, 2015). Due to the recent European crisis numerous companies
face stronger difficulties into achieving good financial performance
and in investing resources to promote innovation (Filippetti and
Archibugi, 2011; Filippetti et al., 2013). Crises acts as a
disruptor of economic activities, and innovation may be considered
both as a privileged dimension of the policy response to them (OECD,
2009) and also one of the most significant drivers of
competitiveness. For that reason an analysis of the impact produced
by innovation in time of crisis may provide interesting additional
insight to literature. Accordingly, we formulate the following
research question (RQ):
\begin{itemize}
\item  Do innovation initiatives, even promoted in time of crisis, produce heterogeneous financial
performance? 
\end{itemize}

\section{Data}

The analysis is performed over the companies listed in the STAR
Market which included, as of 31 December, 2010, 71 companies of
mid-size in terms of capitalization value (between 40 million and 1
billion euros). However, to be consistent, banks and insurance
institutes have been removed from the collected sample, hence
leading to 62 companies. Data have been manually collected from the
consolidated section of the annual reports of the companies, and
taken from the companies  websites. The spanned period is 2006-2010.

Biennium 2006-2007 is the reference pre-crisis period for assessing
the level of innovation of the companies. Innovation is measured by
using two types of indicators, i.e.: the level of tangible and
intangible fixed assets. As discussed in the introduction, these
indicators do not strictly limit "innovation" to "new technology
implementation" (like through buying patents). We consider
"innovation" in a more general sense. Specifically, tangible assets
are intended as the aggregation of the balance sheet items: plants,
machineries and equipments, while we have excluded properties, whose
variations are not necessarily associated to innovation; intangible
assets are obtained by summing items like development costs,
patents, trademarks, licences and concessions, while goodwill is not
taken into account: it can be driven by mergers or acquisition of
new companies.

The triennium 2008-2010 is the time-span -- at time of crisis --
related to the performance. Such performances are measured through
three growth (or decay)  variations, i.e. sales variations, total
assets variations,  and number of employees variations; two
profitability indicators, i.e. Return on investment (ROI) and Return
on sales (ROS); two efficiency indicators, i.e. assets turnover and
sales per employee.

Without loosing too much information and to gain empirical
tractability of the problem, data have been properly treated.
Specifically, in a first study, innovation and performance variables
have been averaged over the reference period.

\subsection{Notations}\label{sec:notations}
The following notations have been adopted.
\begin{itemize}
\item  TIAXyy represents the level of total intangible assets (excluding goodwill) in year
20yy;
\item TTAyy is the level of total tangible assets (excluding properties)
in year 20yy;
\item DSalyy stands for sales variations in year 20yy
\item DAssyy is total assets variations, in year 20yy
\item DLabyy means employees variations, in year 20yy
\item ROIyy is the ROI in year 20yy
\item ROSyy is the ROS in year 20yy
\item ATOyy represents asset turnover, in year 20yy
\item S/Eyy stands for sales per employee, in year 20yy
\end{itemize}
The first two items are those related to innovation, while the
remaining seven ones rely on performance.

Innovation terms are averaged over the period 2006-2007, while the
others in the triennium 2008-2010. 

The adopted notations are:

\begin{itemize}
\item  $<$TIAX$>_2$ is the average total intangible
asset (excluding goodwill) over 2 years: [2006-2007];
\item     $<$TTA$>_2$ represents the average of the total tangible
assets (excluding properties) over 2 years: [2006-2007];
\item     $<$DSal$>_3$ is the average of the sales variations
over 3 years: [2008-2010];
\item     $<$DAss$>_3$ is the average of the total  assets
variations over 3 years: [2008-2010];
\item     $<$DLab$>_30$ represents the average of the number of employees
variations over 3 years: [2008-2010];
\item      $<$ROI$>_3$ is the averaged ROI over 3 years:
[2008-2010];
\item     $<$ROS$>_3$ represents the averaged ROS over 3 years:
[2008-2010];
\item     $<$ATO$>_3$ is the average of the asset turnovers
over 3 years: [2008-2010];
\item     $<$S/E$>_3$ represents the averaged sales per
employee over 3 years: [2008-2010].
\end{itemize}

\section{Methodology} 

The clustering procedure we implement is based on the Voronoi
tessellation, with an asymmetric generalization of the Euclidean
distance. We adapt such methodology to our specific setting.

The final target we have is to compare the companies with respect to
the clusters where they are collected. The clustering procedure is
implemented twice: one for the innovation variables, averaged over
the biennium [2006-2007], and the other for the performance
variables, averaged over the triennium [2008-2010].

In order to avoid scale effects and to be consistent, the variables
of interest have been normalized in the respect of their range of
variation. Formally, for each company $j=1, \dots, 62$, we define:
\begin{equation}
\label{xj} \bar{x}_j=\frac{x_j-m_x}{M_x-m_x},
\end{equation}
where $x$ is the averaged quantity of interest among the nine
innovation and performance variables, the quantity $x_j$ represents
the value of the variable $x$ for the $j$-th company and
$$
m_x=\min\limits_{j=1, \dots, 62} x_j,\qquad M_x=\max\limits_{j=1,
\dots, 62} x_j.
$$
The clustering procedures are then applied to the set of innovation
variables $\mathcal{I}$ and to the set $\mathcal{P}$ collecting the
remaining variables, which are the performance ones. All the
variables are normalized, according to (\ref{xj}) and averaged over
the reference period.

The centroids of the Voronoi tessellation are positive numbers and
will be denoted as $\{\phi_h\}_{h=1}^H$ and $\{\psi_k\}_{k=1}^K$,
where $H$ and $K$ are opportunely chosen integers, for the case of
innovation and performance variables, respectively. The ranges of
variation of the centroids depends on the selected distance measure,
as we will see soon.

We now introduce the weighted Euclidean distance used for the
proposed generalized Voronoi tessellation. Specifically, for the
innovation variables we define:
\begin{equation}
\label{dalpha} d_{\mathcal{I}}(j,\phi_h)=\sum_{x \in \mathcal{I}}
\alpha_x (\bar{x}_j-\phi_h)^2,
\end{equation}
for each centroid $\phi_h$ and where the $\alpha$'s  are the
nonnegative weights of the norm, so that
$$
\sum_{x \in \mathcal{I}} \alpha_x =1.
$$
Analogously, for the performance variables we define:
\begin{equation}
\label{dbeta} d_{\mathcal{P}}(j,\psi_k)=\sum_{x \in \mathcal{P}}
\beta_x (\bar{x}_j-\psi_h)^2,
\end{equation}
for each centroid $\psi_k$ and
$$
\sum_{x \in \mathcal{P}} \beta_x =1.
$$
By definition, we have that $0 \leq d_{\mathcal{I}}(j,\phi_h),
d_{\mathcal{P}}(j,\psi_k) \leq 1$, for each company $j$ and centroid
$\phi_h$ and $\psi_k$.
\newline
The generic Voronoi cell is denoted as $V'_h$ and $V''_k$ for
innovation and performance variables, respectively, where:
$$
V'_h=\{j=1,\dots,
62\,|\,d_{\mathcal{I}}(j,\phi_h)<d_{\mathcal{I}}(j,\phi_{\bar{h}}),\,\,\forall\,\bar{h}
\not=h\};
$$
$$
V''_k=\{j=1,\dots,
62\,|\,d_{\mathcal{P}}(j,\psi_k)<d_{\mathcal{P}}(j,\psi_{\bar{k}}),\,\,\forall\,\bar{k}
\not=k\}.
$$
Of course, the interiors of $V'$s are disjoint sets, so as the
$V''$ 's. Moreover,
$$\bigcup_{h=1}^HV'_h=\bigcup_{k=1}^KV''_k=\{1,\dots,62\}.$$

\subsection{Specifications of the cluster analysis}
As a premise, it is important to point out that the cardinality of
the Voronoi regions might change as the centroids do. Furthermore,
the belonging of the $j$-th company to specific regions provides
information on the level of innovation and on the performance of
$j$.

We have implemented the Voronoi cluster analysis under different
scenarios. For comparison purposes, we have always set $H=K$. The
analyzed cases are now listed:
\begin{itemize}
\item[$I$] $H=K=4$, $\{\phi_h\}_{h=1}^H=\{\psi_k\}_{k=1}^K=\{1/5,2/5,3/5,4/5\}$.
Only one of the $\alpha$''s and $\beta$'s is one, while the values of
the others terms are null. In this case, we explore the clustering
of the companies on the basis of all the individual variables.
\item[$II$] $H=K=4$, $\{\phi_h\}_{h=1}^H=\{\psi_k\}_{k=1}^K=\{1/5,2/5,3/5,4/5\}$, $\alpha_x=1/2$ for each $x \in \mathcal{I}$ and
$\beta_x=1/7$ for each $x \in \mathcal{P}$. This is a \emph{uniform
in value case}, where the definition of the centroids is made by
considering a uniform decomposition of the interval $[0,1]$ and all
the variables are assumed to equally concur in the Voronoi distance.
\item[$III$] $H=K=4$, $\{\phi_h\}_{h=1}^H=\{\psi_k\}_{k=1}^K=\{1/5,2/5,3/5,4/5\}$, $\alpha_x=1/2$ for each $x \in \mathcal{I}$
and the same weight for the macro-variables of the performance
(i.e.: 1/3 for growth, profitability and productivity) with a
uniform distribution of the weights among the variables identifying
each macro-variable. Hence, since growth has three variables of
interest, we assign the same value of the $\beta$''s(i.e.: $1/3
\times 1/3$), while the $\beta$'s for the variables in profitability
and productivity share the same value of $1/3 \times 1/2$. This is a
\emph{uniform in role case}, where the definition of the centroids
is made by considering a uniform decomposition of the interval
$[0,1]$ and all the macro-variables for performance are assumed to
equally concur in the Voronoi distance, with also equal weight for
the variables identifying the three macro-variables for performance.

\end{itemize}
 Scenario $I$ has led to the identification of some outliers,
whose effect is to collapse the majority of the companies in the
first cluster. To remove this inconsistency, 9 companies have been
removed from the sample and the cluster analysis have been applied
to the remaining 53 companies, with data normalized according to
(\ref{xj}).

\section{Results and discussion}

 As we will see below, our analysis leads to a positive
answer to research question RQ. In fact, the overlapping of
the clusters provided through the three clustering methods suggests
the presence of a non straightforward relation between innovation
and performance, even when innovation is performed in a period of
economic crisis. 

Table 1 illustrates the main statistical indicators of the
considered variables.

\begin{center}
INSERT TABLE 1 ABOUT HERE
\end{center}

The distributions of tangible and intangible assets show similar
characteristics: the standard deviation is remarkably higher than
the mean, and skewness and kurtosis show that data are not normally
distributed (with positive skewness and kurtosis).  This
shows that the analyzed companies may have chosen to adopt different
innovative strategies, even if they belong to the same market
segment. The same characteristics also relate to all the growth
indicators, whereas profitability measures tend to a normal
distribution since the median and the mean are substantially similar
and skewness and kurtosis tend to zero. Within the efficiency
measures, the asset turnover and sales per employee are differently
distributed: the former is roughly normally distributed, while the
latter shows a positive skewness and a significant peak.

In Table 2 the distribution of companies among the clusters is shown
either for clustering $II$ and for clustering $III$. To provide
comments on the results, we denote by \textit{first} cluster the one
associated to the smaller centroid and, in an increasing way, the
\textit{second} and the \textit{third} cluster, so that the
\textit{fourth} cluster is the one associated to the higher value of
centroid.

\begin{center}
INSERT TABLE 2 ABOUT HERE
\end{center}

In clustering $II$ the same weight is assigned to all the innovation
and performance indicators, whereas in the clustering $III$ the same
weight is assigned to the perspectives employed for innovation and
performance (i.e. growth, profitability and efficiency).

For what concern innovation, clusterings $II$ and $III$ do not
produce any difference, since in both cases the same weight is
assigned to innovation measures. The greatest number of companies
are located in the first cluster -- i.e., we recall it, the one
associated to the centroid with the lower value -- and a limited
number of companies lie in the second and third clusters. Comparing
clusterings $II$ and $III$, the only difference is that two
companies are reallocated from the first to the third cluster,
meaning that the different weights assigned to growth, profitability
and efficiency, as measures of performance, slightly emphasize
variability. It is  worth noting that no companies are located in the
fourth cluster neither for innovation or performance.

In Table 3, a qualitative description of the clusters of the sample
companies is provided.

\begin{center}
INSERT TABLE 3 ABOUT HERE
\end{center}

The values refer to the clusterings $II$ and $III$ for innovation
and performance. Referring to innovation cluster, the greatest
number of companies (45 out of 53) is located in the first cluster,
meaning that, in relative terms, companies undertake weak innovation
initiatives (at least those which produce reflections on tangible
and intangible assets). Total Assets, total sales and number of
employees -- which are the measures largely employed in literature
for company size -- show that the higher the intensity of
innovation, the higher the size. This is particularly true for total
sales and number of employees. Also the incidence of both tangible
and intangible assets (as percentage of the total assets) is
increasing in the three innovation clusters, meaning that in highly
innovative companies, tangible and intangible assets represent a
relevant portion of the total assets disclosed. The mean/std. dev.
ratio shows that the composition of the clusters is rather
heterogeneous except for the 3rd innovation cluster which is
composed by companies whose size is fairly concentrated around the
mean. For what concern performance, the distribution of companies
among the clusters is quite different from that of innovation. This
provides evidence that the association between innovation and
performance is not self-evident. The averages in the performance
clusters also do not allow to appraise significant differences
neither in terms of company size or incidence of tangible and
intangible assets.

Table 4 shows the averages drivers of innovation and performance for
the entire sample and referred to clustering $II$ and $III$ for
innovation and performance.

\begin{center}
INSERT TABLE 4 ABOUT HERE
\end{center}

Looking at the innovation clustering, a comparison between the three
clusters shows that, reasonably, innovation averages increase from
the first to the third cluster, whereas performance averages show
quite ambiguous tendencies. In the first cluster, the averages of
innovation for tangible and intangible assets are below the general
averages referred to the entire sample, whereas all the performance
indicators are above the general averages. In the second cluster, a
general under-the-general-average performance is associated to an
above-the-general-average innovation. In the third cluster, the
performance averages are mixed. 

The mean/std. dev. ($\mu/\sigma$) ratio allows additional insights about the
homogeneity within the clusters which is generally really low,
meaning that as result of the clustering technique, extremely
different companies lie within the same cluster both in terms of
innovation and performance. The only exception is represented by
asset turnover, since the std. dev. is remarkably concentrated
around the average. This could be interpreted as a possible
association between innovation and asset turnover, even if its
direction remains unclear, since a high asset turnover is associated
to a low innovation in the first cluster, whereas a low asset
turnover is associated to a medium innovation in the second cluster
and then again high asset turnover is associated to high innovation
in the third cluster. It is   worth noting that in the third cluster,
companies appear rather homogeneous in terms of performance,
particularly for profitability (both ROI and ROS) and efficiency
(asset turnover and sales per employee). We could then argue that,
above a particular threshold of innovation intensity, the level
performances seems rather homogeneous, even if  it is  not sure if high
innovation lead to high performance. Similar considerations can be
made for performance clustering. Both in the $II$ and $III$
performance clustering, the performance averages gradually increase
from the first to the third cluster, whereas innovation averages
decrease (intangible assets) or fluctuate (tangible assets). The
relation innovation-performance seems, then, quite puzzling. Even
for performance clustering, heterogeneity generally occurs within
the clusters except for asset turnover.

Some additional considerations may arise by considering the
distances between the average values referred to the performance
clusters. The indicators of growth (particularly sales variation and
employee variation), show substantially an equal distance between
the clusters, whereas in profitability the distance between the
second and third clusters is lower than that between the second and
the first ones, meaning that companies in the third cluster show
performances substantially similar to companies lying in the second
cluster, while an higher distance occurs for companies in the first
cluster.   This could probably mean that there is a low-medium
innovative investment threshold that companies should overcome in
order to get an increase in performance.

In order to provide a synthetic picture, data are consolidated by
industry and the histograms show the distributions among the four
clusters, within each industry. No substantial differences may be
appraised for tangible and intangible assets, whose distributions
are concentrated on the first two clusters. Differently,
profitability shows high frequencies in the second and third
clusters for almost all the industries. Sales per employee is the
only measure of performance that shows distributions similar to
those of tangible and intangible assets. This provides a slight
evidence that sales per employee, among the performance indicators
adopted in the study, is perhaps the most suitable measure to
demonstrate the effects of innovation initiatives on performance.

\section{Conclusions}\label{sec:conclusions}

This paper deals with the exploration of the relationship between
innovation activities  and firms  performance at time of crisis. The
considered sample is given by the companies listed in the STAR
market; the reference period is the quinquennium 2006-2010. Two
innovation and seven performance variables have been manually
collected from the consolidated section of the companies annual
reports.

The analysis is carried out by adopting cluster methodologies based
on Voronoi tessellation. In so doing, the present paper fills an
existing gap between complex science -- with specific reference to
cluster analysis through Voronoi diagrams -- and the field of
microeconomics -- with peculiar attention to the relationship
between innovation at time of crisis and performance. In particular,
three different clustering strategies have been implemented and
discussed.

 Several previous contributions in this field have considered
the effects of innovation strategies on performance using mainly
cross-section data, often in the context of a specific sector
(manufacturing or ICT, for example). The impact produced by
innovation on performance in times of crisis has been rather
neglected. However, our results support what literature basically
asserts for non-crisis periods, i.e.: the performance of the
companies may be sensibly heterogeneous when companies innovate.
\newline
Another important contribution of the present study is related to
the employment of different performance indicators that highlight
the various perspectives, not always convergent, of the business
management. This allowed us to identify the sales per employee
indicator as one of the most suitable measure to intercept the
effects of innovation initiatives on performance, at least in our
sample. 

On the one hand, the motivations for such an heterogeneity remain
unexplored. On the other hand, the findings suggest to continue on
elaborating and arguing the relationship question by employing a
Voronoi tessellation method. Of course, to sort out the various
correlations, causes and effects, represents a very complex task.
Working in this direction should allow better grasp on managing
policies.

\clearpage
 

 \begin{table} \begin{center}
\begin{tabular}[t]{|c| c|c|c|c|c|c| c|c|c|} 
  \hline
 \multicolumn{3}{|c|}{   }   &       \multicolumn{7}{|c|}{   Performance}\\\hline 
    &       \multicolumn{2}{|c|}{   Innovation}       &       \multicolumn{3}{|c|}{   Growth}  &       \multicolumn{2}{|c|}{   Profitability}&     \multicolumn{2}{|c|}{  Efficiency}\\\hline 
 &   Intangible  	&	Tangible   	&	Sales  	&	Tot.Ass. 	&	N.	empl.&	 	&	 	&	Asset & Sales  \\
  &  	&	  	&	 Var.n 	&	  Var.n &  Var.n 	&	ROI	&	ROS	&  & per\\
   &  Assets	&	 Assets	&	  (\%)	&	   (\%)	&    (\%)	&	 	&  	& turnover	& empl.\\\hline
mean ($\mu$)&12,360.46 	&	29,215.40 	& 6\%	&	9\%	&	6\%	&	5\%	&	5\%	&	0,91 	&	275.77 	\\
std.dev.($\sigma$)&18,695.11 	&	45,379.80 	&	14\%	&	16\%	&	14\%	&	5\%	&	7\%	&	0,34 	&	231.20 	\\
$\mu/\sigma$&0.66 	&	0.64 	&	0,46 	&	0,57 	&	0,44 	&	0,85 	&	0,75 	&	2,68 	&	1.19 	\\
  min.&  180.00 	&	86.50 	&	-19\%	&	-10\%	&	-21\%	&	-8\%	&	-14\%	&	0,15 	&	57.20 	\\
Max&80,816.00 	&	  217,237.50 	&	59\%	&	53\%	&	60\%	&	21\%	&	24\%	&	2,04 	&	1,100.76 	\\
Q1&1,346.50 	&	    3,579.50 	&	-1\%	&	-2\%	&	-1\%	&	2\%	&	1\%	&	0,75 	&	148.02 	\\
median&  3,584.00 	&	  10,329.00 	&	3\%	&	4\%	&	3\%	&	4\%	&	5\%	&	0,86 	&	188.04 	\\
Q3&13,917.00 	&	31,331.50 	&	12\%	&	16\%	&	11\%	&	8\%	&	9\%	&	1,09 	&	281.07 	\\
skewness&2.28 	&	2.57 	&	1.48 	&	1.32 	&	1.41 	&	0.44 	&	0.27 	&	0.78 	&	2.25 	\\
 kurtosis&      5.14 	&	           6.79 	&	3.60 	&	1.07 	&	4.13 	&	0.60 	&	0.76 	&	1.76 	&	5.05 	\\ \hline
\end{tabular}
 \caption{  Main statistical indicators of the innovation and performance variables.; $Tot.Ass.$ = "Total Assets"; $Var.n$ = "variation"; $empl.$ = "employee"; $N.$ = "Number of".} \label{Table1}
\end{center} \end{table}
\clearpage

 \begin{table} \begin{center}
\begin{tabular}[t]{|c| c|c|c|c|c|c| } 
  \hline
 \multicolumn{7}{|c|}{  II clustering  }   \\  \hline
 \multicolumn{2}{|c}{} &   \multicolumn{5}{|c|}{  Performance }     \\ \hline
	&	 	&	1st cluster	&	2nd cluster	&	3rd cluster	&	4th cluster	&	Tot	 \\ \hline	
Innovation	&	1st cluster	&	16	&	22	&	7	&	0	&	45	 	\\	
	&	2nd cluster	&	2	&	2	&	0	&	0	&	4	 \\	
	&	3rd cluster	&	1	&	3	&	0	&	0	&	4	 	\\	
	&	4th cluster	&	0	&	0	&	0	&	0	&	0 	\\	 \hline
	&	Tot	&	19	&	27	&	7	&	0	&	53	  \\ \hline	
   \hline	
   \multicolumn{7}{|c|}{  III clustering  }\\ \hline
 \multicolumn{2}{|c}{} &   \multicolumn{5}{|c|}{  Performance }     \\ \hline
	&		&	 	1st cluster	&	2nd cluster	&	3rd cluster	&	4th cluster	&	Tot.	\\ \hline	
Innovation	&	 	1st cluster		&	14	&	22	&	9	&	0	&	45	\\	
	 	&		 2nd cluster	&	2	&	2	&	0	&	0	&	4	\\	
 		&		 3rd cluster	&	1	&	3	&	0	&	0	&	4	\\	
	 	&		 4th cluster	&	0	&	0	&	0	&	0	&	0	\\	\hline
		 &		 Tot			&	17	&	27	&	9	&	0	&	53	 	 \\ \hline	
\end{tabular}
 \caption{  Distribution of companies among the clusters, either for clustering $II$ or for clustering $III$ } \label{Table2}
\end{center} \end{table}

  \clearpage

   \begin{table} \begin{center}
\begin{tabular}[t]{|c|c| cccccc|} 
  \hline 
   \multicolumn{2}{|c|}{} &	&	$Tot.Ass.$ 	&	Total Sales 	&	$N.$	&	\% Intangible	&	\% Tangible	\\
 \multicolumn{2}{|c|}{Nr. of} 	&		&	2006-2007	&	2006-2007	&	$empl.s$ 	&	 Assets on 	&	 Assets on 	\\
 \multicolumn{2}{|c|}{Companies	}&		&	 (\EUR/1.000)	&	 (\euro/1.000)	&	 2006-2007	&	 $Tot.Ass.$ &  $Tot.Ass.$ 	\\ \hline
	&	 	53	&	Mean	&	303,053 	&	267,689 	&	1,324 	&	5\%	&	10\%	\\
	&	 	&		Std. Dev. 	&	304,144 	&	261,828 	&	1,534 	&	6\%	&	12\%	\\ \hline \hline
    \multicolumn{7}{|c}{Innovation II/III clustering} \\ \hline
	&	 	 	&	Mean	&	241,199 	&	210,452 	&	945 	&	4\%	&	8\%	\\
 	1st cluster 	&	45	&	Std. Dev. 	&	190,099 	&	187,814 	&	883 	&	5\%	&	8\%	\\
 		&	&	Mean/St. Dev.	&	1.27 	&	1.12 	&	1.07 	&	0.88 	&	1.04 	\\ \hline
 &	 	&	Mean	&	731,745 	&	467,442 	&	2,974 	&	8\%	&	16\%	\\
 2nd cluster	&	4	&	Std. Dev. 	&	798,924 	&	469,023 	&	3,205 	&	8\%	&	22\%	\\
 &		&	Mean/St. Dev.	&	0.92 	&	1.00 	&	0.93 	&	0.97 	&	0.73 	\\ \hline
 &	 	&	Mean	&	570,226 	&	711,853 	&	3,936 	&	12\%	&	22\%	\\
 3rd cluster	&	4	&	Std. Dev. 	&	261,414 	&	329,233 	&	2,232 	&	9\%	&	27\%	\\
	&		&		Mean/St. Dev.	&	2.18 	&	2.16 	&	1.76 	&	1.39 	&	0.82 	\\ \hline\hline
	
    \multicolumn{7}{|c}{Performance II clustering} \\ \hline
    &		 &	Mean	&	210,607 	&	157,375 	&	828 	&	8\%	&	10\%	\\  
 1st cluster 	&	19	&	Std. Dev. 	&	130,310 	&	118,218 	&	584 	&	7\%	&	11\%	\\  
	&		&	Mean/St. Dev.	&	1.62 	&	1.33 	&	1.42 	&	1.14 	&	0.94  	\\ \hline
&	 	&	Mean	&	385,628 	&	354,434 	&	1,885 	&	3\%	&	9\%	\\
 2nd cluster	&	27	&	Std. Dev. 	&	387,615 	&	293,834 	&	1,959 	&	3\%	&	13\%	\\
	&		&		Mean/St. Dev.	&	0.99 	&	1.21 	&	0.96 	&	0.90 	&	0.71  	\\ \hline
	&		 	&	Mean	&	235,476 	&	232,525 	&	503 	&	3\%	&	10\%	\\
	 3rd cluster	&	7	&	Std. Dev. 	&	228,112 	&	340,089 	&	396 	&	5\%	&	12\%	\\
	&		&		Mean/St. Dev.	&	1.03 	&	0.68 	&	1.27 	&	0.62 	&	0.85 	 	\\ \hline \hline
	 
     \multicolumn{7}{|c}{Performance III clustering} \\ \hline
     &		 &	Mean	&	212,193 	&	143,475 	&	770 	&	8\%	&	9\%	\\ 
 1st cluster 	&	17	&	Std. Dev. 	&	130,561 	&	106,315 	&	500 	&	7\%	&	11\%	\\
 &	& Mean/St. Dev.	&	1.63 	&	1.35 	&	1.54 	&	1.11 	&	0.80 	 	\\ \hline
 &	 	&	Mean	&	379,176 	&	351,761 	&	1,948 	&	3\%	&	10\%	\\
2nd cluster	&	27	&	Std. Dev. 	&	391,123 	&	296,006 	&	1,940 	&	4\%	&	13\%	\\
&		&	Mean/St. Dev.	&	0.97 	&	1.19 	&	1.00 	&	0.78 	&	0.79 	\ 	\\ \hline
	&	 	&	Mean	&	246,310 	&	250,098 	&	497 	&	3\%	&	9\%	\\
3rd cluster	&	9	&	Std. Dev. 	&	202,108 	&	299,683 	&	357 	&	4\%	&	11\%	\\
	&		&		Mean/St. Dev.	&	1.22 	&	0.83 	&	1.39 	&	0.71 	&	0.84 	\\ \hline
\end{tabular}
 \caption{ Statistical description of the clusters for the sample companies;.  $Tot.Ass.$ = "Total Assets"; $empl.$ = "employees"; $N.$ = "Number of"} \label{Table3}
\end{center} \end{table}
  \clearpage

   \begin{table} \begin{center}
\begin{tabular}[t]{|cc| ccccccccc|} 
  \hline 
&    \multicolumn{2}{c}{}  Intangible &	Tangible 	&	Sales 	&	Tot.Ass. &	Empl. &	ROI	&	ROS	&	Asset  	&	Sales\\	
 &       \multicolumn{2}{c}{} Assets	&	 Assets	&	 \% Var.n	&	 \% Var.n	&	 \% Var.n	&	 	&		& turn.r	&	/empl.	\\ \hline \hline
Ent. &	Mean ($\mu$)	&	  12,360 	&	  29,215 	&	6\%	&	9\%	&	6\%	&	5\%	&	5\%	&	0.91 	&	275.77 	\\	
 &	Std.Dev.($\sigma$)	&	  18,695	&	  45,380 	&	14\%	&	16\%	&	14\%	&	5\%	&	7\%	&	0.34 	&	231.20 	\\	
 	&	$\mu/\sigma$	&	0.66 	&	0.64 	&	0.46 	&	0.57 	&	    0.44 	&	    0.85 	&	   0.75 	&	  2.68 	&	   1.19 	\\	\hline\hline
    \multicolumn{7}{|c}{Innovation II/III clustering} \\ \hline
1st &Mean ($\mu$)	&	7,127	&	18,554 	&	7\%	&	9\%	&	7\%	&	5\%	&	6\%	&	0.93 	&	293.83 	\\	
cl.	&	Std.Dev.($\sigma$)	&	  9,311 	&	24,023 	&	15\%	&	17\%	&	15\%	&	6\%	&	7\%	&	0.36 	&	248.53	\\	
	&	$\mu/\sigma$	 &	0.77 	&	0.77 	&	0.49 	&	0.55 	&	0.46 	&	      0.87 	&	     0.79 	&	2.57 	&	1.18 	\\	
2nd  &	Mean ($\mu$)	&	28,081 	&	71,512 	&	4\%	&	3\%	&	1\%	&	2\%	&	2\%	&	0.67 	&	165.90 	\\	
cl.	&	Std.Dev.($\sigma$)	&	29,505 	&	65,192 	&	11\%	&	7\%	&	2\%	&	6\%	&	10\%	&	0.13 	&	58.91	\\	
	&	$\mu/\sigma$	 &	0.95 	&	1.10 	&	0.34	&	0.35	&	0.47	&	0.25	&	0.16	&	5.36 	&	2.81	\\	
3rd  &	Mean ($\mu$)		&	55,511 	&	106,862 	&	1\%	&	14\%	&	1\%	&	4\%	&	5\%	&	0.97 	&	182.49 	\\	
cl.	&	Std.Dev.($\sigma$)	&	28,454 	&	107,418 	&	5\%	&	18\%	&	5\%	&	1\%	&	2\%	&	0.20 	&	48.84	\\	
	&	$\mu/\sigma$	 &	1.95 	&	0.99 	&	0.19	&	0.81	&	0.21	&	3.00	&	2.57	&	4.91 	&	3.73	\\	\hline\hline
    \multicolumn{7}{|c}{Performance II clustering} \\ \hline	
1st   &	Mean ($\mu$)		&	16,356 	&	 22,484 	&	-2\%	&	-4\%	&	-2\%	&	0\%	&	0\%	&	0.79 	&	222.75 	\\	
	cl.&	Std.Dev.($\sigma$)	&	 20,762 	&	31,548 	&	10\%	&	4\%	&	8\%	&	4\%	&	7\%	&	0.28 	&	157.65	\\	
	&	$\mu/\sigma$		&	0.79 	&	0.71 	&	0.21 	&	0.83 	&	0.24 	&	0.05 	&	0.05 	&	2.85 	&	1.41 	\\	
2nd  &	Mean ($\mu$)		&	 11,930 	&	  35,841 	&	9\%	&	12\%	&	8\%	&	7\%	&	8\%	&	0.94 	&	262.81 	\\	
	cl.&	Std.Dev.($\sigma$)	&	 19,346 	&	56,259 	&	9\%	&	13\%	&	11\%	&	4\%	&	5\%	&	0.25 	&	232.07	\\	
	&	$\mu/\sigma$	 	&	0.62 	&	0.64 	&	0.94	&	0.92	&	0.71	&	1.54	&	1.46	&	3.77 	&	1.13	\\	
3rd  &	Mean ($\mu$)		&	3,174 	&	 21,931 	&	20\%	&	35\%	&	20\%	&	9\%	&	12\%	&	1.11 	&	469.70 	\\	
cl.	&	Std.Dev.($\sigma$)	&	   4,744 	&	 32,962 	&	24\%	&	17\%	&	21\%	&	6\%	&	9\%	&	0.65 	&	332.70	\\	
	&	$\mu/\sigma$		&	0.67 	&	0.67 	&	0.84	&	2.02	&	0.93	&	1.49	&	1.26	&	1.71 	&	1.41	\\	 \hline\hline
     \multicolumn{7}{|c}{Performance III clustering} \\ \hline
1st  &	Mean ($\mu$)		&	 17,515 	&	  22,521 	&	-1\%	&	-4\%	&	-1\%	&	0\%	&	-1\%	&	0.79 	&	197.21 	\\	
cl.	&	Std.Dev.($\sigma$)&	 21,701 	&	32,921 	&	9\%	&	4\%	&	7\%	&	3\%	&	6\%	&	0.26 	&	102.36	\\	
	&	$\mu/\sigma$		&	   0.81 	&	0.68 	&	0.09 	&	0.83 	&	0.20 	&	0.09 	&	0.11 	&	3.00 	&	1.93 	\\	
2nd  &	Mean ($\mu$)		&	  11,944 	&	36,171 	&	8\%	&	11\%	&	8\%	&	6\%	&	8\%	&	0.91 	&	221.52 	\\	
cl.	&	Std.Dev.($\sigma$)	&	  19,349 	&	56,146 	&	11\%	&	13\%	&	11\%	&	4\%	&	6\%	&	0.26 	&	137.49	\\	
	&	$\mu/\sigma$	 &	     0.62 	&	     0.64 	&	0.71	&	0.88	&	0.69	&	1.46	&	1.37	&	3.54 	&	1.61	\\	
3rd &	Mean ($\mu$)		&	  3,873 	&	20,993 	&	16\%	&	27\%	&	14\%	&	9\%	&	11\%	&	1.14 	&	586.90 	\\	
cl.	&	Std.Dev.($\sigma$)	&	  4,337 	&	29,926 	&	22\%	&	22\%	&	22\%	&	5\%	&	8\%	&	0.57 	&	373.88	\\	
	&	$\mu/\sigma$		&	0.89 	&	0.70 	&	0.73	&	1.23	&	0.62	&	1.73	&	1.34	&	2.02 	&	1.57	\\	 \hline
\end{tabular}	 \caption{ Main statistical characteristics of the Innovation and Performance  variables for the whole sample (Ent.) and inside the clusters (cl.); $Tot.Ass. $ = Total assets; $Var.n$ = variation;  $turn.r$= turnover; $empl.$= employee} \label{Table4}
\end{center} \end{table}

\end{document}